\documentclass[useAMS]{mn2e}
\usepackage{graphicx}
\usepackage{pslatex}

\title[Chemical abundance analysis of HD~76431]{Spectral analysis and abundances of the post-HB star HD 76431}
\author[V. Khalack et al.]
{V. Khalack$^{1}$, B. Yameogo$^{1}$, F. LeBlanc$^{1}$, G. Fontaine$^{2}$,  E. Green$^{3}$, V. Van Grootel$^{4}$, P. Petit$^{5,6}$ \\
$^{1}$D\'{e}partement de physique et d'astronomie, Universit\'{e} de Moncton, 18 avenue Antonine-Maillet, Moncton, N.-B., Canada E1A 3E9  \\
$^{2}$D\'{e}partement de physique, Universit\'{e} de Montr\'{e}al, C.P. 6128, Succursale Centre-Ville, Montr\'{e}al, QC, Canada H3C 3J7 \\
$^{3}$Steward Observatory, University of Arizona, 933 North Cherry Avenue, Tucson, AZ 85721, USA \\
$^{4}$Institut d'Astrophysique et de G\'{e}ophysique, Universit\'{e} de Liege, All\'{e}e du 6 Aout 17, 4000 Liege, Belgium  \\
$^{5}$Universit\'{e} de Toulouse, UPS-OMP, Institut de Recherche en Astrophysique et Plan\'{e}tologie, Toulouse, France \\
$^{6}$CNRS, Institut de Recherche en Astrophysique et Plan\'{e}tologie, 14 Avenue Edouard Belin, F-31400 Toulouse, France
}
\begin{document}

\date{}
\pagerange{\pageref{firstpage}--\pageref{lastpage}} \pubyear{2014}

\maketitle

\label{firstpage}

\begin{abstract}
HD~76431 is a slow rotating post-HB star that shows an underabundance of helium by 0.5 dex relative to the solar value.
These observational facts suggest that atomic diffusion could be active in its atmosphere.
We have used the MMT and Bok spectra to estimate the atmospheric parameters of the target star
using the model atmospheres and synthetic spectra calculated with TLUSTY and SYNSPEC.
The derived values of the effective temperature, surface gravity, helium abundance are consistent with those obtained by Ramspeck et al. \shortcite{ram01b}.
It appears that NLTE effect are not important for HD~76431.
We have used Stokes I spectra from ESPaDOnS at CFHT to perform an abundance analysis and a search for
observational evidence of vertical stratification of the abundance of certain elements.
The results of our abundance analysis are in good agreement with previously published data with respect to average abundances.
Our numerical simulations show that carbon and nitrogen reveal signatures of vertical abundance stratification in the atmosphere of HD~76431.
It appears that the carbon abundance increases toward the deeper atmospheric layers.
Nitrogen also shows a similar behaviour, but in deeper atmospheric layers we obtain a significant dispersion for the estimates of its abundance.
To our knowledge, this is the first demonstration of vertical abundance stratification of metals in a post-HB star and up to now it is the hottest star
to show such stratification features.
We also report the detection of two Si\,{\sc iii} and one Ti\,{\sc iii} emission lines in the spectra of HD~76431 that were not detected in previous studies.
\end{abstract}


\section{Introduction}

Horizontal-branch (hereafter HB) stars have evolved past the main sequence and burn helium in their core which is surrounded by a hydrogen burning shell.
In general, after core helium exhaustion, they evolve towards the asymptotic giant branch (AGB), but a part of HB stars does not reach the AGB stage and is usually called the extreme horizontal branch (EHB) \cite{Dorman+93}. The boundary between EHB stars that evolve mainly to hotter temperatures and those that evolve towards the AGB is near $T_{\rm eff}$ = 20000~K on the zero age extended horizontal branch (ZAEHB) \cite{Dorman+93}.  The hottest stars, those with the thinnest hydrogen envelopes, evolve to higher temperatures during and after core helium burning and completely bypass the AGB.  Stars with effective temperatures near 20000~K on the ZAEHB still have very small envelopes, which are nevertheless sufficiently thick to allow the star to evolve towards the AGB for a while after core helium exhaustion, although the shell burning is soon quenched and the star contracts again to hotter temperatures \cite{Ostensen+12}.

HB stars with effective temperatures larger than approximately 11500~K are of particular interest since they exhibit abundance anomalies (Glaspey et al. 1989; Behr et al. 1999; Moehler et al. 1999; Behr, Cohen \& McCarthy 2000; Behr 2003a) such as under-abundances of helium and over-abundances of several metals including iron.
HB stars with $T_{\rm eff}$ above the 11500~K  threshold also show low rotational velocities as compared to cooler HB stars. The rotational velocity of these hot HB stars drops to a value of $V$sin$i\simeq$ 10 km s$^{-1}$ or less (Peterson, Rood \& Crocker 1995; Behr et al. 2000a,b; Behr 2003b; Recio-Blanco et al. 2004). Such a drop in the rotational velocity is thought to lead to a more hydrodynamically stable atmosphere where atomic diffusion (Michaud 1970) may take place. Queivy et al. (2009) demonstrated that for HB stars with such low rotational velocities, the helium convection zone disappears because meridional circulation is not strong enough to prevent helium from settling gravitationally. The atmosphere therefore becomes stable and atomic diffusion leads to vertical abundance stratifications and detectable surface abundance anomalies.

Other observational anomalies are detected due to the presence of vertical abundance stratification in the atmospheres of these blue HB stars. For example, a photometric jump in the ($u,u-y$) colour-magnitude diagram is observed at $T_{\rm eff}\simeq$ 11500~K in several globular clusters (Grundahl et al. 1999). Photometric gaps are also detected at this $T_{\rm eff}$ (Ferraro et al. 1998). These two photometric anomalies were theoretically confirmed by the model atmospheres of Hui-Bon-Hoa, LeBlanc \& Hauschildt (2000) and LeBlanc et al. (2009). These models include the effect of the vertical stratification of the elements on the atmospheric structure which can explain the observed photometric jumps and gaps (LeBlanc, Hui-Bon-Hoa \& Khalack 2010).

Khalack et al. (2007, 2008 and 2010) detected vertical stratification of certain elements including iron in several blue HB stars. The stars studied there are found in the $T_{\rm eff}$ = 10750 to 15500~K  range. These results serve as additional proof that atomic diffusion is at play in their atmosphere.

The lower $T_{\rm eff}$ limit where abundance stratification in HB stars occurs is relatively well established at approximately 11500~K. However, the upper limit in $T_{\rm eff}$ where no such stratification exists is not as well established. The results of Moni Bidin et al. (2012) for HB stars in $\omega$ Centauri shows that helium is underabundant for stars up to approximately 32000~K. It suggests that this might give the upper limit where other physical processes such as mass loss for instance, could dominate over atomic diffusion.

This paper aims to verify if vertical stratification of the elements is present in the post-HB star HD~76431. This is by far the hottest star ($T_{\rm eff}$ = 31000~K; Ramspeck, Heber \& Edelmann 2001) for which a detailed abundance analysis that verifies for the presence of vertical stratification has been undertaken. The results from this spectral analysis could give insight on whether or not atomic diffusion is still dominant in such hot stars.

\section{Details concerning HD~76431}

HD~76431 was found to be evolved past the HB phase by Ramspeck et al. \shortcite{ram01b} (see their Figure 5). This has also been confirmed by the results of Chountonov \& Geier \shortcite{cho12} (see their Figure 1).

\subsection{Observations and data reduction}
\label{obs}

Our analysis is based on high-resolution spectropolarimetric observations carried out with ESPaDOnS at CFHT\footnote{The Canada-France-Hawaii Telescope (CFHT) is operated by the National Research Council of Canada, the Institut National des Sciences de l'Univers of the Centre National de la Recherche Scientifique of France, and the University of Hawaii.} \cite{Petit+12}. Seventeen spectra were obtained in the range of 3700\AA\, to 10000\AA\, at a spectral resolution of 65000 with aim to search for the signatures of magnetic field \cite{otool+05}. Petit et al. \shortcite{Petit+12} have confirmed the results of Elkin \shortcite{Elkin98} and Chountonov \& Geier \shortcite{cho12}, and found no detectable Zeeman signatures in the Stokes I and V spectra of HD~76431.

\begin{figure}
\includegraphics[width=3.3in,angle=-90]{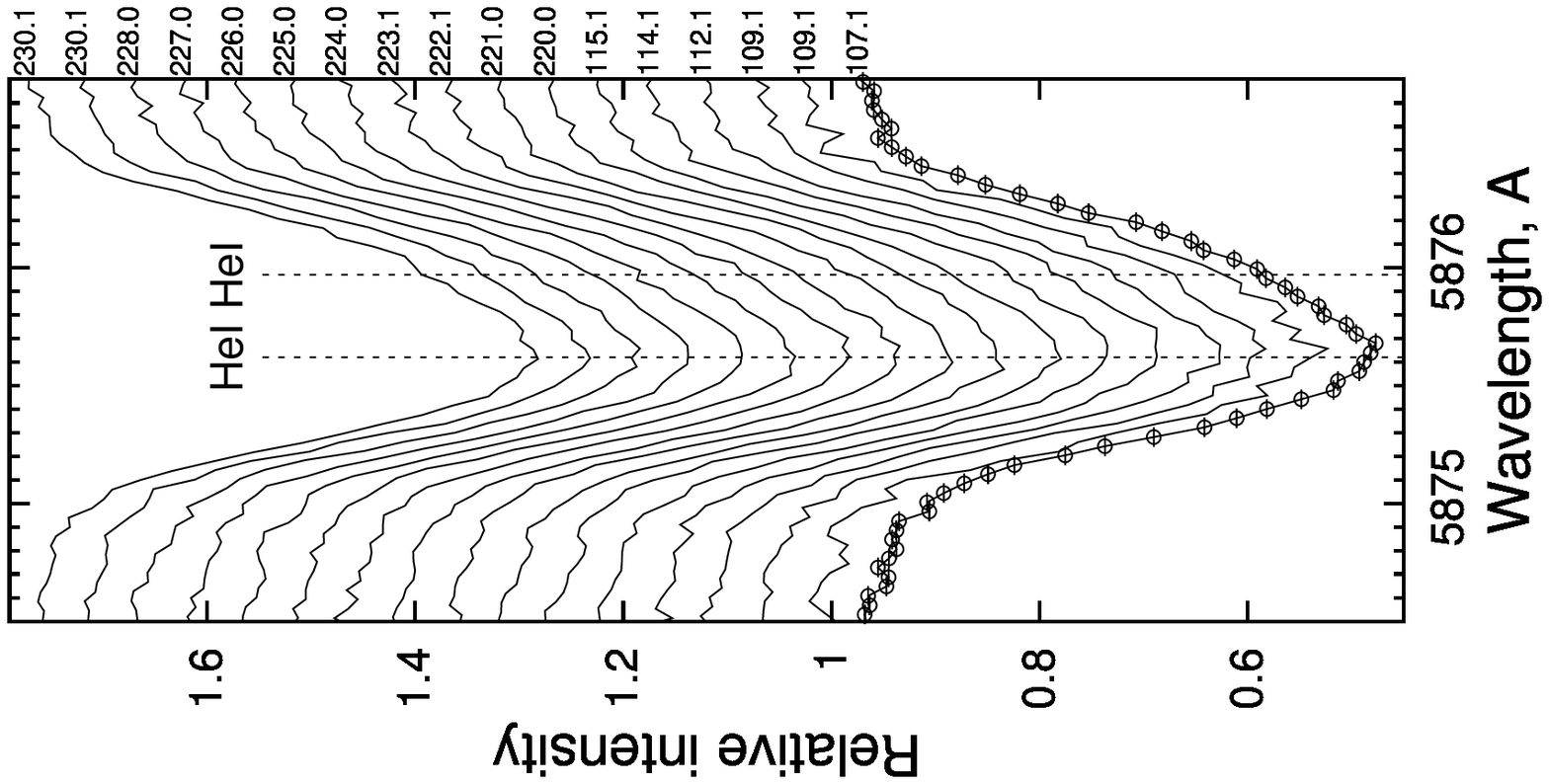}
\includegraphics[width=3.3in,angle=-90]{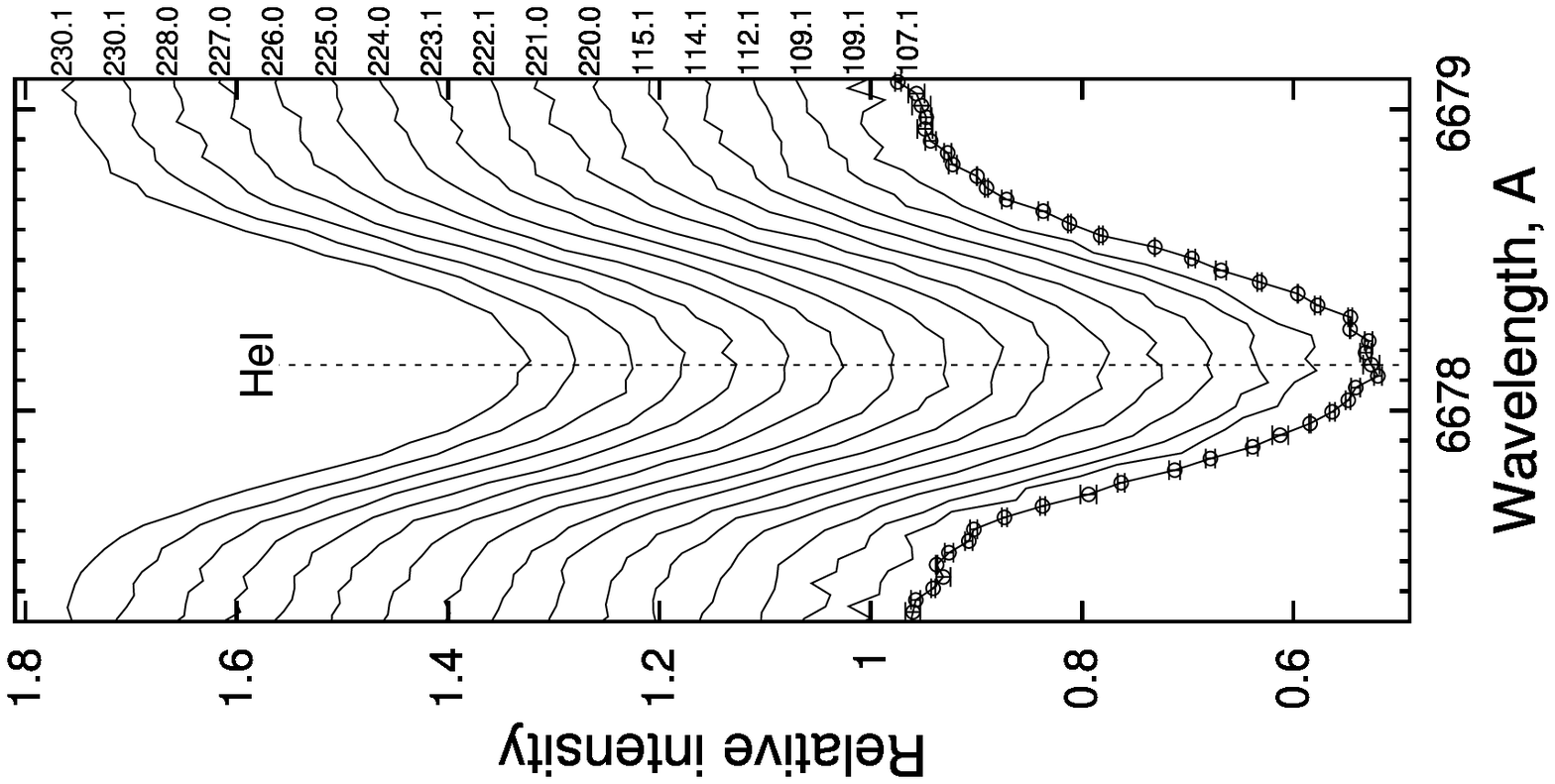}
\caption{ Profiles of He\,{\sc i} 5875\AA\, (right panel) and He\,{\sc i} 6678\AA\, (left panel) absorption lines obtained during the different dates of observations. The spectra are shifted vertically by 0.05 for better visibility. On the right side of each panel the time of each observation is presented with respect to the HJD=2455000. For the first observation with HJD=2455107.1 we show the observational errors that have almost the same value for the other spectra presented here. }
\label{fig1}
\end{figure}

A detailed pre-analysis of the 17 spectra obtained by Petit et al. \shortcite{Petit+12} has shown that the profiles of almost all the visible spectral lines do not vary much with
the date of observation (see for example Fig.~\ref{fig1}) spanning from 2009 Oct. 02 to 2010 Feb. 02 (see Petit et al. 2012).
This fact does not argue in favour that HD~76431 might be in a close binary system \cite{cho12}.
Taking into account the detected stability of the line profiles, we have composed all these spectra into a single spectrum which has been used for the abundance analysis presented here.

\begin{figure*}
\includegraphics[width=2.45in,angle=-90]{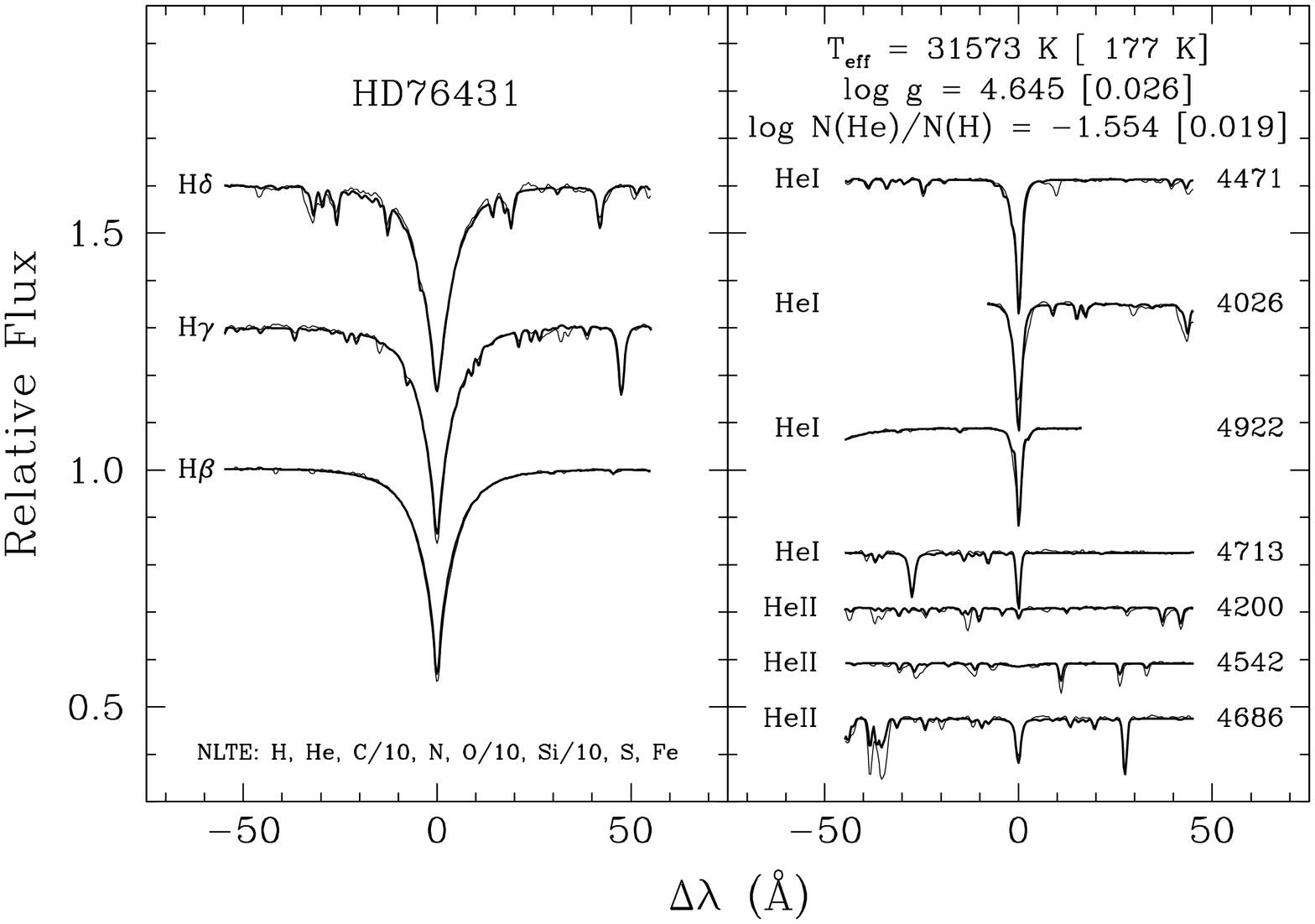}
\includegraphics[width=2.45in,angle=-90]{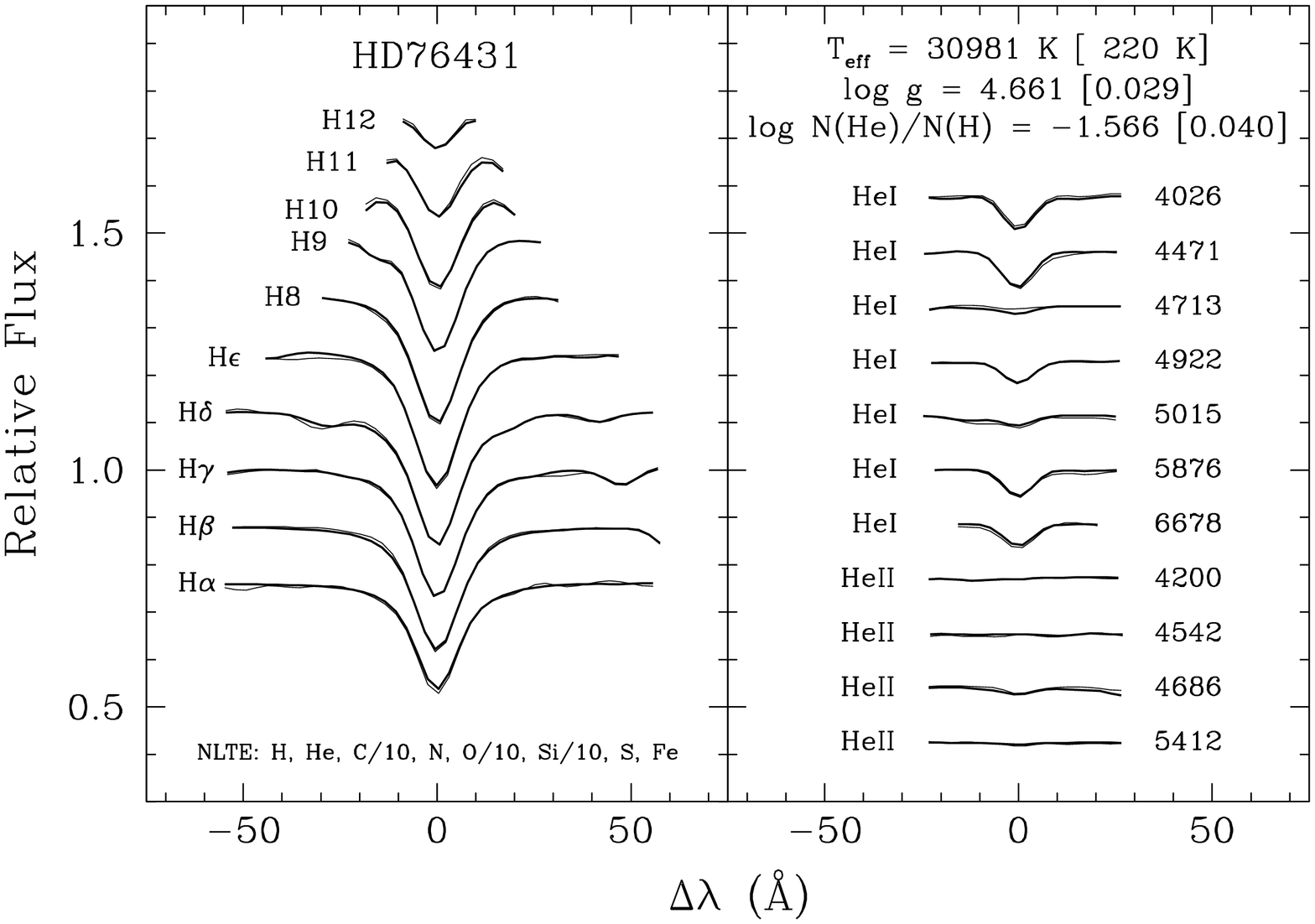}
\caption{The effective temperature and gravity derived from the fitting Balmer, He\,{\sc i} and He\,{\sc ii} line profiles in the MMT (left) and Bok spectra (right) of HD~76431.}
\label{fig1b}
\end{figure*}

Nine low-resolution spectra of HD 76431 were obtained with the B\&C Cassegrain spectrograph on Steward Observatory’s 2.3 m Bok telescope on Kitt Peak
between 1999 and 2010. The 400/mm first-order grating was used with a 2.5 arcsec slit to obtain spectra with a typical resolution of 9\AA\, (R$\sim$560)
over the wavelength interval 3620 - 6900 \AA\AA. The instrument rotator was set prior to each exposure to align the slit within $\sim$2$\degr$
of the parallactic angle at the midpoint of the exposure. Eight intermediate-resolution spectra of HD~76431 were taken with the Blue spectrograph
on the 6.5 m MMT on Mount Hopkins, Arizona between 1996 and 1998. The 832/mm second-order grating and 1.0 arcsec slit gave a resolution of 1.05\AA\,(R$\sim$4200)
over 4000 - 4950 \AA\AA. Again, the slit was always aligned with the parallactic angle. Exposure times on both telescopes were chosen to achieve S/N of 100 - 200
for each of the individual spectra.

The Bok and MMT spectra were bias-subtracted, flat-fielded, background-subtracted, optimally extracted, wavelength-calibrated and flux calibrated
using standard IRAF\footnote{IRAF is distributed by the National Optical Astronomy Observatories, which are operated by the Association of Universities for Research in Astronomy, Inc., under cooperative agreement with the National Science Foundation.} tasks (Tody 1986; 1993).
Each set was combined with median filtering to derive very high S/N, time-averaged spectra with S/N of 465 and 525, respectively.

\subsection{Effective temperature and the surface gravity}
\label{Teff}

Ramspeck et al. \shortcite{ram01b} have derived for this star $T_{\rm eff}$ = 31000~K and log $g$ = 4.51 from the analysis of the Balmer and helium lines.
In order to check these values, we carried out an analysis aimed at obtaining independent estimates of the atmospheric parameters of the target star
on the basis of our Steward Observatory spectra. For that purpose, we used an available grid of NLTE, fully metal-blanketed model atmospheres and synthetic spectra.
That grid was computed with the public codes TLUSTY and SYNSPEC run in parallel mode on the cluster CALYS at Universit\'{e} de Montr\'{e}al currently made up of 320 fast processors. Details on TLUSTY and SYNSPEC may be found on TLUSTY's Website\footnote{http://nova.astro.umd.edu} and in Lanz \& Hubeny (2003; 2007).

The chosen metallicity is derived from the work of Blanchette et al. \shortcite{Blanchette+08} who analyzed in detail the FUSE spectra of five hot subdwarf stars.
For the most abundant metals, these authors found typical abundances in those objects of 0.1 times the solar values for C, O, and Si,
and nearly solar values for N, S, and Fe. We adopted this metallicity as representative of the class of Extreme Horizontal Branch Stars in this checking exercise.
Note therefore that this is used only in the evaluation of the fundamental parameters of HD~76431,
which, as we show below, are not sensitive to variations in metal content in the domain of effective temperature$-$surface gravity of interest.
Our models include the following ions: H\,{\sc i} and H\,{\sc ii}, C\,{\sc ii} to C\,{\sc v}, N\,{\sc ii} to N\,{\sc vi},
O\,{\sc ii} to O\,{\sc vii}, Si\,{\sc ii} to Si\,{\sc v}, S\,{\sc ii} to S\,{\sc vii}, and Fe\,{\sc ii} to Fe\,{\sc viii}.
The grid includes models with $T_{\rm eff}$ between 20000~K and 50000~K in steps of 2000~K, log $g$ values between 4.4 and 6.4 in steps of 0.2~dex,
and $\log{N_{\rm He}/N_{\rm H}}$ values between $-$4.0 and 0.0 in steps of 0.5~dex.

For a given observational spectrum, we computed suitable synthetic spectra by degrading the resolution to the experimental value
(1.0\AA\, for the MMT spectrum and 8.7\AA\, for the Bok spectrum).
All of the available hydrogen and helium lines in the spectrum were then simultaneously fitted using a $\chi^2$ minimization procedure in 3D
similar to that of Saffer et al. \shortcite{Saffer+94}. More details on this approach may be found in Latour et al. (2011; 2013).

Figure~\ref{fig1b} shows the results of our fitting procedure. The quality of the fits is excellent and, moreover, both spectra lead to consistent results.
A close examination of the fit of the MMT spectrum reveals, not surprisingly, that the metal content is not the same in HD~76541 as in our models with
a representative metallicity. This is not very significant, however, in terms of establishing the fundamental parameters of the star.
Indeed, by redoing the same exercise but using, this time, a metal-free grid, we find the following results:
$T_{\rm eff}$ = 31390$\pm$370~K, log $g$ = 4.70$\pm$0.05, and $\log{N_{\rm He}/N_{\rm H}}$ = $-$1.57$\pm$0.03 for the MMT spectrum,
and $T_{\rm eff}$ = 31180$\pm$220 K, log $g$ = 4.67$\pm$0.03, and $\log{N_{\rm He}/N_{\rm H}}$ = $-$1.58$\pm$0.05 for the Bok spectrum.
These values are, within the uncertainties, essentially the same as those shown in Figure~\ref{fig1b}, which demonstrates
that the exact metallicity has little bearing on the derived values of the atmospheric parameters in the domain of interest.
We wish to point out that the uncertainties quoted in Figure~\ref{fig1b} are only formal errors of the fit and, consequently,
underestimate the true uncertainties due to systematic bias that can be always present in any grid of calculated models of stellar atmospheres.
Nevertheless, this method provides values of atmospheric parameters that are in good
agreement with the results obtained through the asteroseismology \cite{Green+11}.

\begin{table}
\centering
\caption{\label{tab1} List of spectral lines used for the abundance analysis. The full version of table is available online.}
\begin{tabular}{llllc}
\hline
\hline
Ion & $\lambda$, \AA& $\log{N_{\rm ion}/N_{\rm H}}$& $\log gf$ & $E_{\rm l}$, $cm^{-1}$ \\
\hline
C\,{\sc ii} &  4313.106 & -3.406$\pm$0.025 & -1.1361 &  186443.69  \\
C\,{\sc ii} &  5133.282 & -3.554$\pm$0.007 & -1.8218 &  166990.73  \\
C\,{\sc ii} &  5151.085 & -3.576$\pm$0.008 & -1.8218 &  167035.71  \\
C\,{\sc ii} &  5662.460 & -3.549$\pm$0.003 & -1.8218 &  167035.71  \\
C\,{\sc ii} &  5648.070 & -3.514$\pm$0.012 & -1.5932 &  166990.73  \\
C\,{\sc ii} &  6582.880 & -4.255$\pm$0.006 & -2.9631 &  116537.65  \\
C\,{\sc ii} &  6779.940 & -3.608$\pm$0.014 & -2.0503 &  166990.73  \\
\hline
\end{tabular}
\end{table}

In addition, we explicitly verified that NLTE effects are not important in the present case
by repeating again the same fitting procedure with a third grid of models, a metal-free one again, but computed in the LTE approximation.
With this, we now find, $T_{\rm eff}$ = 30440$\pm$250~K, log $g$ = 4.67$\pm$0.04, and $\log{N_{\rm He}/N_{\rm H}}$ = $-$1.59$\pm$0.03 for the MMT spectrum,
and $T_{\rm eff}$ = 30680$\pm$190 K, log $g$ = 4.70$\pm$0.04, and $\log{N_{\rm He}/N_{\rm H}}$ = $-$1.56$\pm$0.05 for the Bok spectrum.
Comparison of these results provides an estimate of the dispersion of the derived values of atmospheric parameters.
We have derived essentially the same values of $T_{\rm eff}$ (within 2$\sigma$), log $g$ (within 1$\sigma$), and $\log{N_{\rm He}/N_{\rm H}}$ (within 1 $\sigma$)
from the analysis of two spectra obtained at different telescopes with different setups, different spectral resolutions, and different spectral coverages.
In view of this, we conclude that our derived fundamental parameters for HD~76431 are in good agreement between themselves, and with those estimated by Ramspeck et al. \shortcite{ram01b}.

\begin{table*}
\centering
\caption{\label{tab2} Average abundances of chemical species in the stellar atmosphere of HD~76431 assuming $T_{\rm eff}$ = 31000~K and log $g$ = 4.5 }
\begin{tabular}{llllc}
\hline
\hline
      &      \multicolumn{4}{c}{Abundance ($\log{N_{\rm ion}/N_{\rm H}}$)} \\
Ion   &     \multicolumn{2}{c}{This paper$^a$}        &  Ramspeck$^a$ &   Solar$^b$      \\
      & $\xi$=0 km s$^{-1}$ & $\xi$=5 km s$^{-1}$ &  et al. (2001) &              \\
\hline
C\,{\sc ii}  &	-3.50$\pm$0.52 (27) & -3.67$\pm$0.54 (19)& -3.51$\pm$0.09 (7) &	-3.61$\pm$0.05 \\
C\,{\sc iii} &	-3.38$\pm$0.15 (22) & -3.44$\pm$0.17 (24)& -3.45$\pm$0.16 (15)&	-3.61$\pm$0.05 \\
N\,{\sc ii}  &	-3.93$\pm$0.19 (53) & -3.90$\pm$0.22 (51)& -3.92$\pm$0.17 (65)&	-4.22$\pm$0.06 \\
N\,{\sc iii} &	-3.86$\pm$0.16 (10) & -3.68$\pm$0.19 (7) & -3.82$\pm$0.09 (10)&	-4.22$\pm$0.06 \\
O\,{\sc ii}  &	-4.16$\pm$0.38 (55) & -4.16$\pm$0.30 (44)& -4.19$\pm$0.19 (44)&	-3.34$\pm$0.05 \\
Ne\,{\sc ii} &	-4.38$\pm$0.41 (5)  & -4.39$\pm$0.42 (5) & -3.70$\pm$0.17 (5) &	-4.16$\pm$0.06 \\
Mg\,{\sc ii} &	-4.54$\pm$0.84 (3)  & -4.71$\pm$0.65 (3) & -4.78$\pm$0.36 (2) &	-4.47$\pm$0.09 \\ 
Al\,{\sc iii}&	-6.08$\pm$0.10 (5)  & -6.21$\pm$0.27 (6) & -6.01$\pm$0.24 (5) &	-5.63$\pm$0.06 \\
Si\,{\sc iii}&	-5.21$\pm$0.41 (7)  & -5.39$\pm$0.59 (5) & -5.05$\pm$0.26 (7) &	-4.49$\pm$0.04 \\
Si\,{\sc iv} &	-5.06$\pm$0.31 (4)  & -5.00$\pm$0.33 (5) & -5.28$\pm$0.15 (5) &	-4.49$\pm$0.04 \\
S\,{\sc iii} &	-5.28$\pm$0.23 (5)  & -5.08$\pm$0.14 (5) & -5.41$\pm$0.08 (6) &	-4.86$\pm$0.05 \\
Ar\,{\sc ii} &	-5.36$\pm$0.11 (5)  & -5.29$\pm$0.09 (6) &  &	-5.82$\pm$0.08 \\
Ti\,{\sc iii}&	-4.34$\pm$0.15 (2)  & -4.29$\pm$0.14 (2) &  & 	-7.10$\pm$0.06 \\
Fe\,{\sc iii}&	-4.95$\pm$0.11 (17) & -4.80$\pm$0.31 (21)& -4.78$\pm$0.27 (16)&	-4.55$\pm$0.05 \\
\noalign{\smallskip}
\hline
\end{tabular}
\\
\scriptsize{ {\it Notes:} $^a$The number in parenthesis represents the number of selected lines \\
$^b$Solar data are taken from Grevesse et al. \shortcite{Grevesse+10} }
\end{table*}

Taking into account that the {\sc Zeeman2} code \cite{Landstreet88} is designed for the fitting of individual line profiles in LTE, the values for the fundamental parameters of HD~76431 derived by Ramspeck et al. \shortcite{ram01b} are employed in the present study. The stellar atmosphere model has been calculated with the PHOENIX code \cite{Hauschildt+99} and used  for our spectral analysis. It should be noted that according to the evolutionary path associated to this star (see Figure 5 of Ramspeck et al. 2001), when it was on the HB its effective temperature was in the range from 20000~K to 25000~K.

Ramspeck et al. \shortcite{ram01b} also found that helium is underabundant by 0.5 dex relative to its solar value and that $V$sin$i$ is less than 5 km s$^{-1}$. Behr (2003b) also determined that its rotational velocity is near zero. The helium abundance and the low rotational velocity argues that diffusion could be active in this star and therefore that vertical abundance stratification could be present in its atmosphere.

\section{Chemical abundances}

\subsection{Method of analysis}

The spectrum synthesis code {\sc Zeeman2} (developed by Landstreet 1988 and modified by Wade et al. 2001)
has been used to perform the line profile simulations.
Khalack \& Wade \shortcite{Khalack+Wade06} have modified this code to
allow for an automatic minimization of the model parameters using the
{\it downhill simplex method} \cite{press+}.

To verify for the presence of vertical abundance stratification of chemical species, we have studied the dependence of the abundance derived from
each analyzed profile relative to the optical depth $\tau_{\rm 5000}$ for the list of selected line profiles. For each analysed ion, we compose a list of line profiles that belong to this ion and are free from blends. This list is presented at the Table~\ref{tab1}, where the first and the second columns contain the name of ion and the wavelength of each line. The following columns give the abundance estimate along with its error, the $\log gf$ value and the energy of the lower level respectively. A statistically significant vertical stratification of the element's abundance can be measured based on the analysis of at least 10 or more line profiles that belong to one or two ions of this element that are clearly visible in the analyzed spectrum \cite{kha08}.

\begin{figure*}
\includegraphics[width=2.3in,angle=-90]{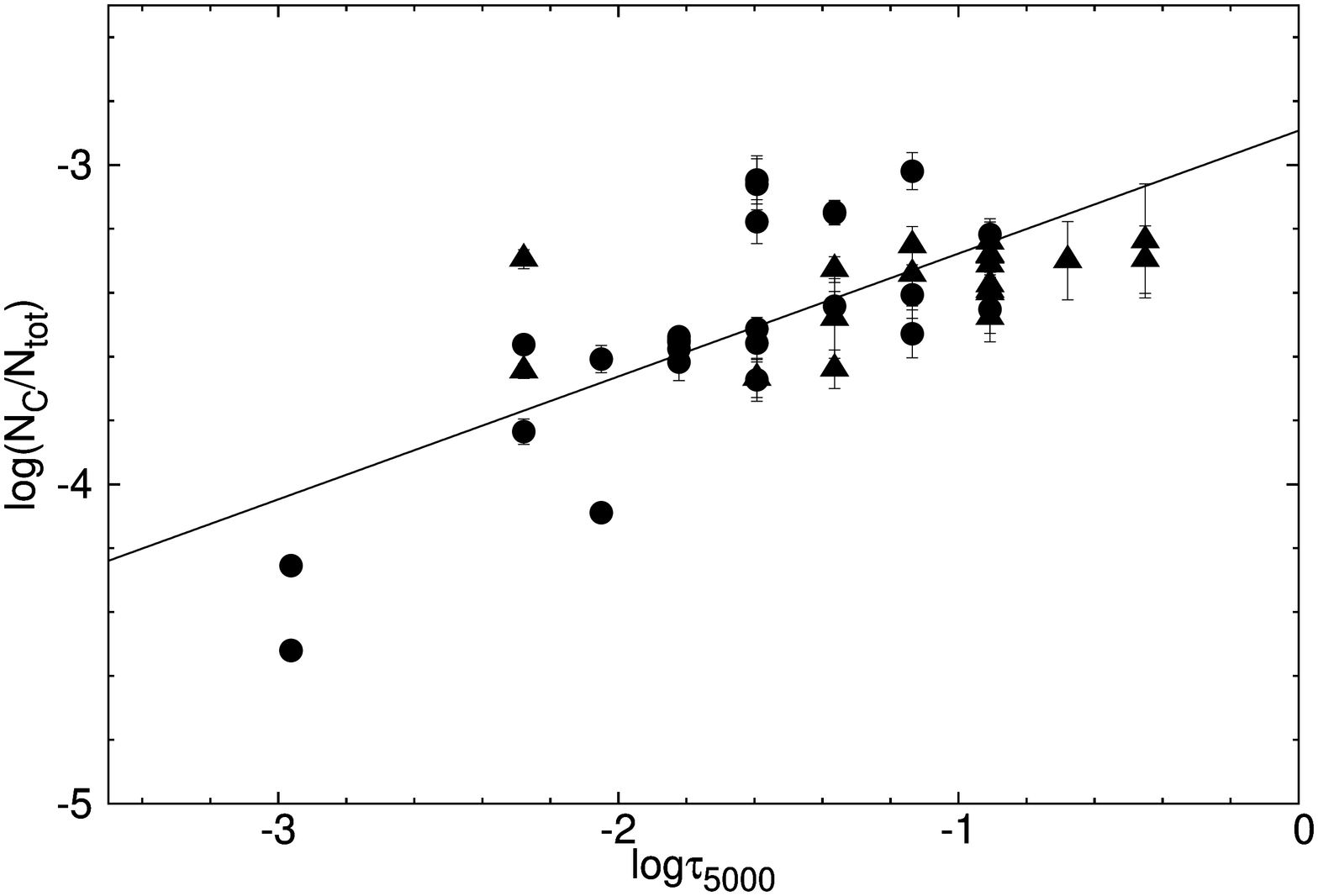}
\includegraphics[width=2.3in,angle=-90]{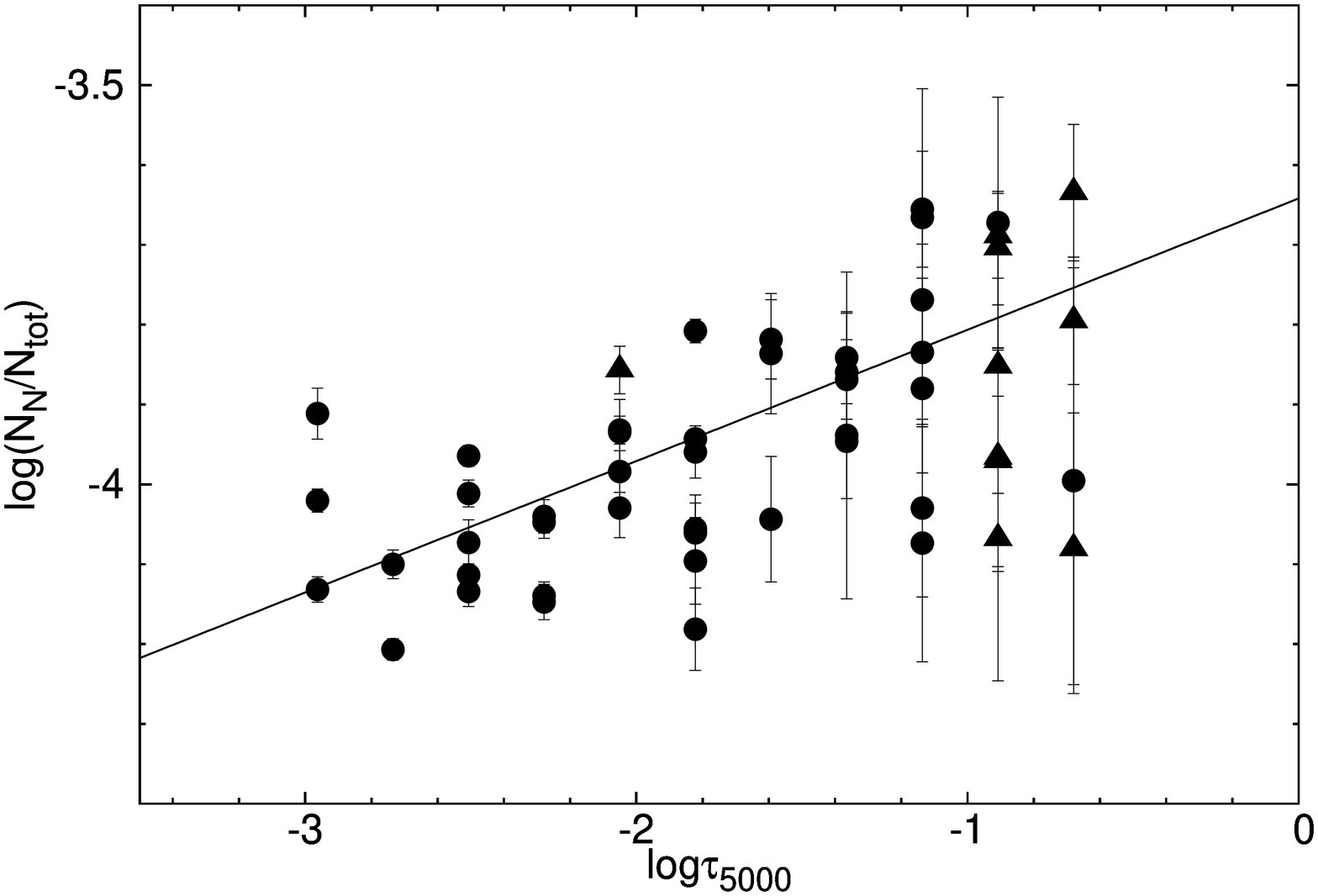}
\caption{ Abundance estimates from the analysis of C\,{\sc ii} (filed circles) and  C\,{\sc iii} (filed triangles) lines on the left panel, and of
N\,{\sc ii} (filed circles) and  N\,{\sc iii} (filed triangles) lines on the right panel as a function of line (core) formation optical depth. }
\label{fig2}
\end{figure*}

\begin{figure*}
\includegraphics[width=2.3in,angle=-90]{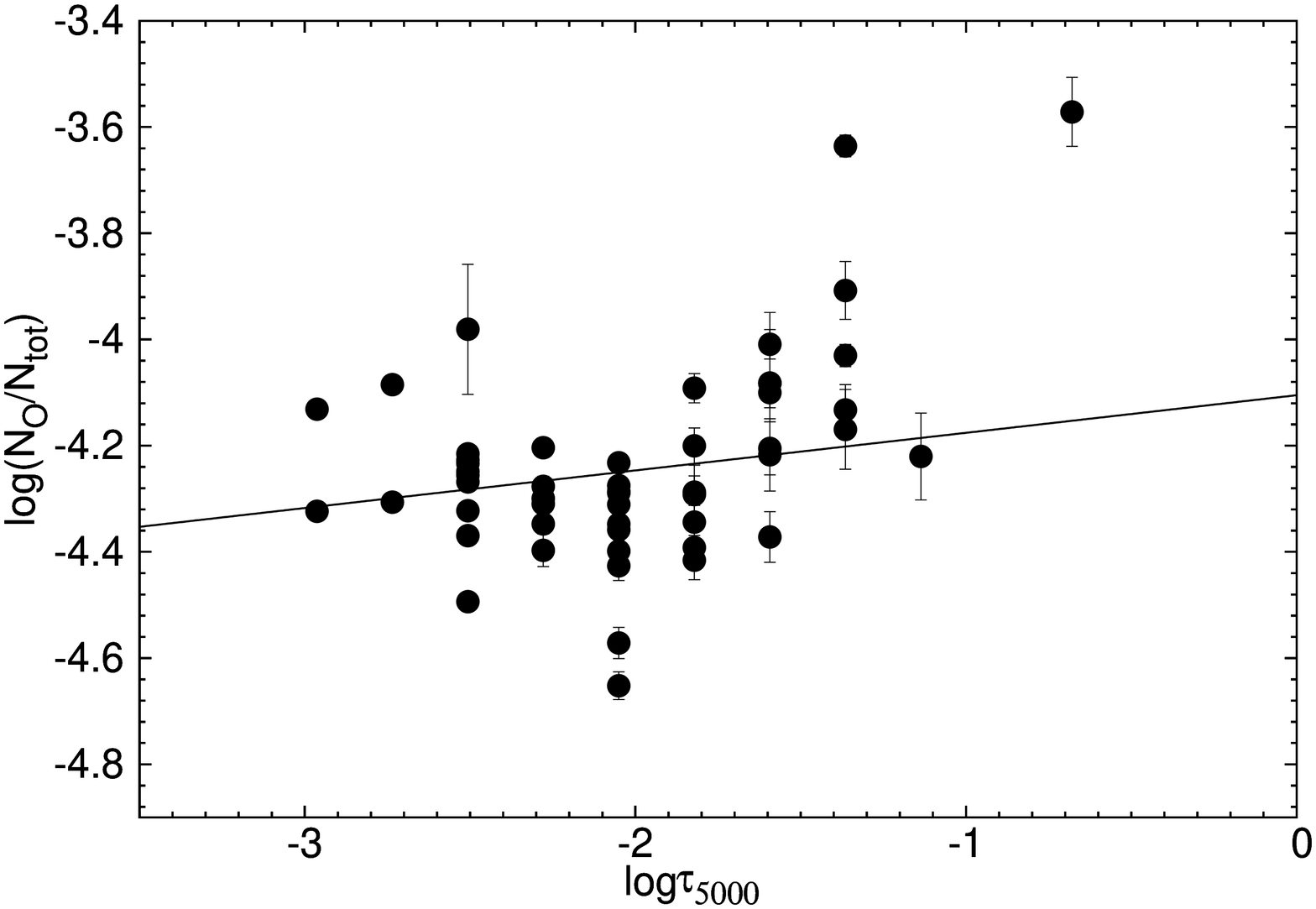}
\includegraphics[width=2.3in,angle=-90]{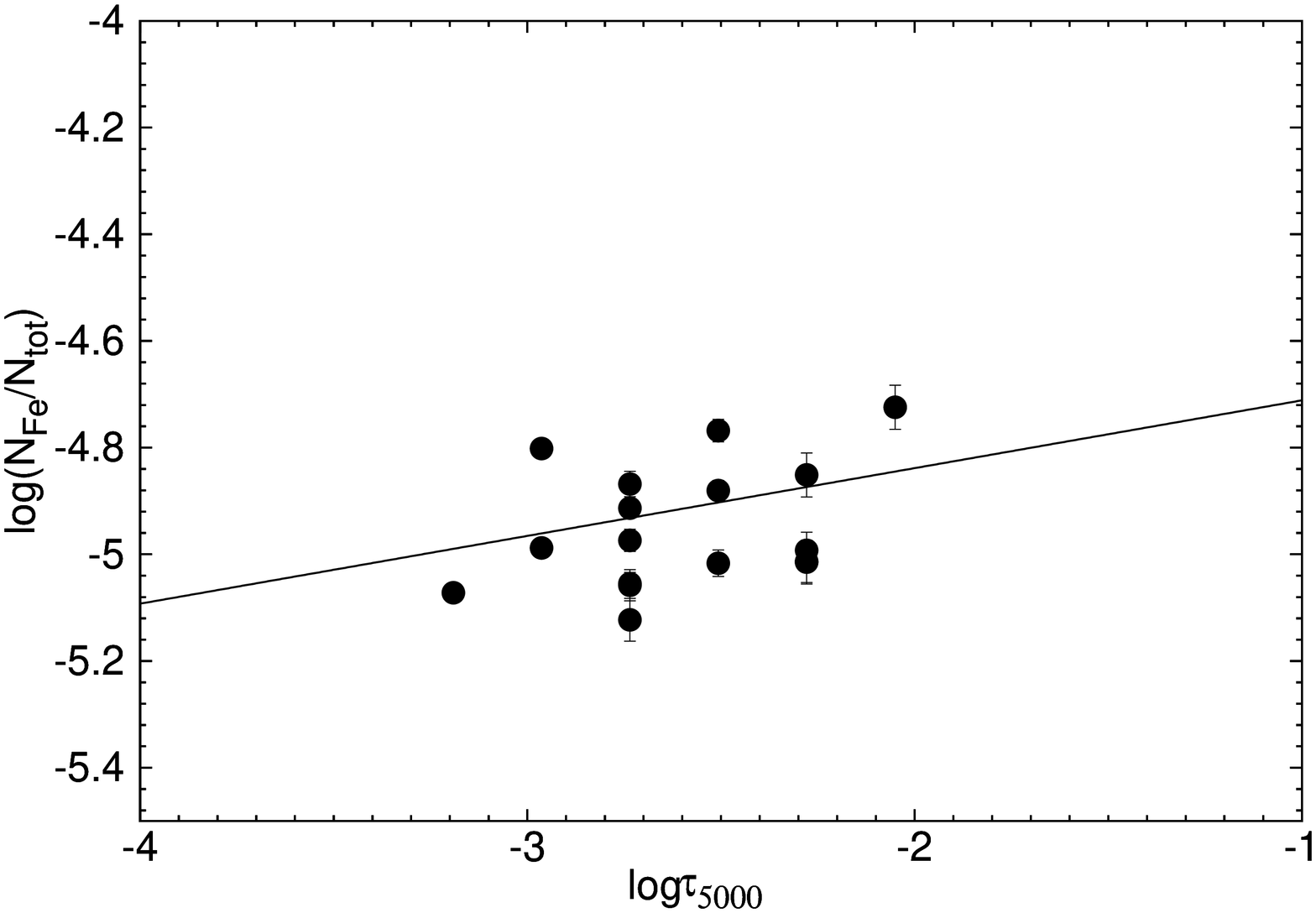}
\caption{Same as Fig.~\ref{fig2}, but from the analysis of O\,{\sc ii} (left panel) and Fe\,{\sc iii} (right panel) lines.}
\label{fig3}
\end{figure*}

We assume that the core of the line profile is formed mainly at line optical depth $\tau_{\rm \ell}$=1, which corresponds to a
particular layer of the stellar atmosphere with a
certain value of the continuum optical depth $\tau_{\rm 5000}$ that in turn corresponds to a given layer of the stellar atmosphere model.
In this way, from the simulation of each line that belongs to a particular ion, we can obtain its abundance
(that corresponds to an optical depth $\tau_{\rm 5000}$), and the value of $V \sin{i}$ and the radial velocity of the star in question.
A comprehensive description of the fitting procedure is given by Khalack et al. \shortcite{kha07}.
Providing that the cores of line profiles are generally formed at different optical depths $\tau_{\rm 5000}$, we can study the vertical distribution of element abundance from the analysis of elements showing at least 10 or more line profiles that belong to the same ion of this element.
The aforementioned method was also used by Khalack et al. \shortcite{kha08,kha10} 
for the study of stratification in BHB stars and by Thiam et al. \shortcite{Thiam+10} for HgMn stars.

The stellar atmosphere model of HD~76431 has been calculated with the {\sc Phoenix} code
\cite{Hauschildt+99} assuming LTE (Local Thermodynamic Equilibrium), and the fundamental parameters $T_{\rm eff}$ = 31000~K, log $g$ = 4.5 as well as
the abundances of chemical species obtained by Ramspeck et al. \shortcite{ram01b} for this star.
The abundances of the remaining elements were kept at their solar values.

\subsection{Average abundances}

Almost all absorption lines in the all 17 observed spectra of HD~76431 are stable and do not show a statistically significant variability with the time of observation.
The observed variability of several lines is caused by the different overlapping with the telluric lines during different nights of observation.
Figure~\ref{fig1} presents an example of He\,{\sc i} 5875\AA\, and 6678\AA\, line profiles that do not show any strong variability during the whole period of observations.
Therefore, we have combined the 17 observed spectra into the one by simply adding them together taking into account the shift of the wavelength scale.
The spectra were already corrected for the Earth orbital motion and the wavelength scales were therefore shifted in various spectra.
The final spectrum was then re-normalised and used for the abundance analysis.

In this study, we have selected a list of spectral lines that are suitable for abundance and stratification analysis that belong to the following ions
C\,{\sc ii}, C\,{\sc iii}, N\,{\sc ii}, N\,{\sc iii}, O\,{\sc ii}, Ne\,{\sc ii}, Mg\,{\sc ii}, Al\,{\sc iii}, Si\,{\sc iii}, Si\,{\sc iv},
S\,{\sc iii}, Ar\,{\sc ii}, 
Ti\,{\sc iii} and Fe\,{\sc iii}.
The full list of the analyzed spectral lines is given in the electronic edition of {\it MNRAS}.
Atomic data for the selected lines were extracted from VALD-2
(Kupka~et~al. 1999, Ryabchikova~et~al. 1999)
and NIST \cite{Kramida+13} line databases.

Table~\ref{tab2} presents the average values of the abundance for each analysed ion.
The first column of the Table~\ref{tab2} gives the name of analysed ion, while the second and the third columns contain the abundance estimates
obtained from the fitting of observed line profiles assuming the microturbulent velocity $\xi$= 0 km s$^{-1}$ and $\xi$= 5 km s$^{-1}$ respectively.
The fourth column presents the abundance estimates for the same ions obtained by Ramspeck et al. \shortcite{ram01b}, and the fifth column gives
their solar abundance \cite{Grevesse+10}. In the brackets on the right of the abundance in the second, third and fourth columns,
the number of analysed spectral lines that belong to the respective ion is given.

Our simulations show that there are no significant difference between the results for average abundances
obtained assuming a microturbulent velocity $\xi$= 0 and 5 km s$^{-1}$. Our results are also in good agreement with the abundances obtained
by Ramspeck et al. \shortcite{ram01b} for HD~76431 who assumed a microturbulent velocity $\xi$= 5 km s$^{-1}$ (see Table~\ref{tab2}).
For example, we have obtained the same abundance (taking into account the error bars) for Mg\,{\sc ii} as in Ramspeck et al. \shortcite{ram01b},
but these values are different from the abundance $\log{N_{\rm Mg}/N_{\rm H}}$ = - 5.13$\pm$0.05 reported by Behr \shortcite{b444} for this star.
The only significant difference one can observe is for Ne\,{\sc ii}. Ramspeck et al. \shortcite{ram01b} have reported that Ne\,{\sc ii} is significantly overabundant in HD~76431,
while our results argue in favor of an abundance close to its solar value. Contrarily to Ramspeck et al. \shortcite{ram01b}, we have not found
any line that belongs to P\,{\sc iii} in the spectra, but have identified lines of Ar\,{\sc ii} 
and Ti\,{\sc iii}.

In this study, each line profile has been fitted individually to obtain independent abundance estimates for the analysed chemical species.
The average abundance value is based on these estimates that can have a relatively large scatter due to the vertical stratification of element abundance with the optical depths (see Section~\ref{CNstrat}) and/or due to the uncertainties in the atomic data (uncertainty of the log $gf$ values, absence of information for damping coefficients, etc.), and the unaccounted small contribution of blends (line profiles with strong blends were excluded from the analysis). The larger scatter of the abundance estimates leads to the larger error bars obtained for the average abundance of chemical species. Meanwhile, the abundances published by Ramspeck et al. \shortcite{ram01b} are derived using the classical curve-of-growth method followed by a detailed spectrum analysis (simultaneous fitting) of all studied line profiles employing the LINFOR program \cite{ram01a}, that results in smaller (compared to this study) error bars for the average abundance estimates.

We have also performed an abundance analysis employing the stellar atmosphere models calculated for log $g$ = 4.51 and $T_{\rm eff}$=33000~K and 29000~K, and for $T_{\rm eff}$=31000~K with log $g$ = 4.75 and 4.25 to verify for the influence of the uncertainties in the determination of $T_{\rm eff}$ and log $g$ (see subsection \ref{Teff}) onto the obtained values for average abundances of chemical species. Our simulations show that the derived abundance of certain elements (Ar\,{\sc ii}, Ne\,{\sc ii}, Si\,{\sc iv}) is sensitive to the value of effective temperature or to both $T_{\rm eff}$ and log $g$ (C\,{\sc iii}), while the abundance of other elements is not significantly affected by the aforementioned variation of the effective temperature and gravity
since the deviations of derived abundances are smaller than the error bars presented in Table~\ref{tab2}. Nevertheless, the change of the effective temperature by 2000~K (assuming the same gravity) or the gravity by 0.25 (assuming the same effective temperature) still results in a statistically significant stratification of carbon and nitrogen abundance with optical depth (see subsection~\ref{CNstrat}).

The totality of the analysed line profiles (see Table~\ref{tab2}) have been used to determine an average value of $V\sin{i}$ = 3.5$\pm$0.5 km s$^{-1}$
and radial velocity $V_{\rm r}$ = 47$\pm$1 km s$^{-1}$ which are in good agreement with the results reported by Ramspeck et al. \shortcite{ram01b}
and by Behr \shortcite{b444}.

\subsection{Stratification of C and N abundances }
\label{CNstrat}

Among the analysed ions, only C\,{\sc ii}, C\,{\sc iii}, N\,{\sc ii}, N\,{\sc iii}, O\,{\sc ii} and Fe\,{\sc iii}
show a sufficient number of lines in the spectrum of HD~76431 to verify if their abundance changes with optical depth.
For each of the aforementioned ions, we have individually analyzed all line profiles that possess small errors
for the estimated parameters (abundance, $V_{\rm r}$ and $V\sin{i}$) assuming zero microturbulence.
The measured abundance for each line was then associated to a line depth formation. The data for C\,{\sc ii}
and C\,{\sc iii}, and for N\,{\sc ii} and N\,{\sc iii} are combined because they are related to the same chemical element.
The elements' abundances obtained for different optical depths (see Figs.~\ref{fig2} and ~\ref{fig3}) were fitted to a linear function
(for the range of atmospheric depths $-3.5 < \log{\tau_{5000}} < 0.0$) using the least-square algorithm to statistically
evaluate the significance of observable trends. This range of $\log{\tau_{5000}}$ was chosen because it includes all the lines of the aforementioned ions.

The main result found here is that the carbon and nitrogen abundances seem to increase towards the deeper atmosphere (see Fig.~\ref{fig2}).
The slope of these gradients represents the increase of $\log{N_{ion}/N_{H}}$ (in dex) calculated per dex of $\log{\tau_{5000}}$.
In the case of carbon and nitrogen, the slopes appear to be statistically significant and reach the values $a_{\rm C}$=0.67$\pm$0.05
and $a_{\rm N}$=0.16$\pm$0.03 respectively. Meanwhile, in the case of oxygen (slope $a_{\rm O}$=0.07$\pm$0.05) and iron (slope $a_{\rm Fe}$=0.13$\pm$0.09)
their abundances remain almost constant for different optical depths (see Fig.~\ref{fig3}).

In order to gauge the possible effect of the uncertainty surrounding the underlying model atmosphere used here, we analysed this star using models with $T_{\rm eff}$ = 33000~K and 29000~K (assuming log $g$ = 4.51), and models with log $g$ = 4.75 and 4.25 (assuming $T_{\rm eff}$ = 31000~K).
Our simulations show that an increase of effective temperature by 2000~K or decrease of gravity by 0.26 do not effect the final results.
Application of stellar atmosphere model with a lower temperature ($T_{\rm eff}$ = 29000~K, log $g$ = 4.51) or with a higher gravity
($T_{\rm eff}$ = 31000~K, log $g$ = 4.75) result in a steeper slope of the vertical stratification of nitrogen abundance with optical depth,
while the results for carbon are not changed much.
This test 
reaffirms the validity of our results about the presence of
vertical stratification of nitrogen and carbon in the atmosphere of HD~76431.

Comparing the behaviour of carbon and nitrogen abundance in Fig.~\ref{fig2} we may conclude that the abundance of carbon increases
strongly and steadily relative to the atmospheric depth, while the increase of nitrogen abundance is smaller and there is more scatter
of the abundance for the deeper atmospheric layers around $\log{\tau_{5000}}$= -1.0.

\subsection{Emission line profiles}

The spectra of HD~76431 show a presence of weak emission lines Si\,{\sc iii} 8102.86\AA, 8103.45\AA\, and Ti\,{\sc iii} 8235.58\AA\, (see Fig.~\ref{fig4}).
The two Si\,{\sc iii} emission lines remain stable during the whole period of observation, while the Ti\,{\sc iii} varies during the first week of observation.
For the HJD=2455114.1 this line appears to have a depression that transforms its profile into a P-Cygni profile.
At the same date, a depression is detected in the left wing of the emission line Si\,{\sc iii} 8102.86\AA.
In both cases these depressions may be caused by telluric lines. Nevertheless, the aforementioned emission lines belong to the star and might be caused by the presence of
an upper stellar atmosphere of lower density composed of highly ionised metals like Si\,{\sc iii} and Ti\,{\sc iii}.

\section{Discussion}

\begin{figure}
\includegraphics[width=3.3in,angle=-90]{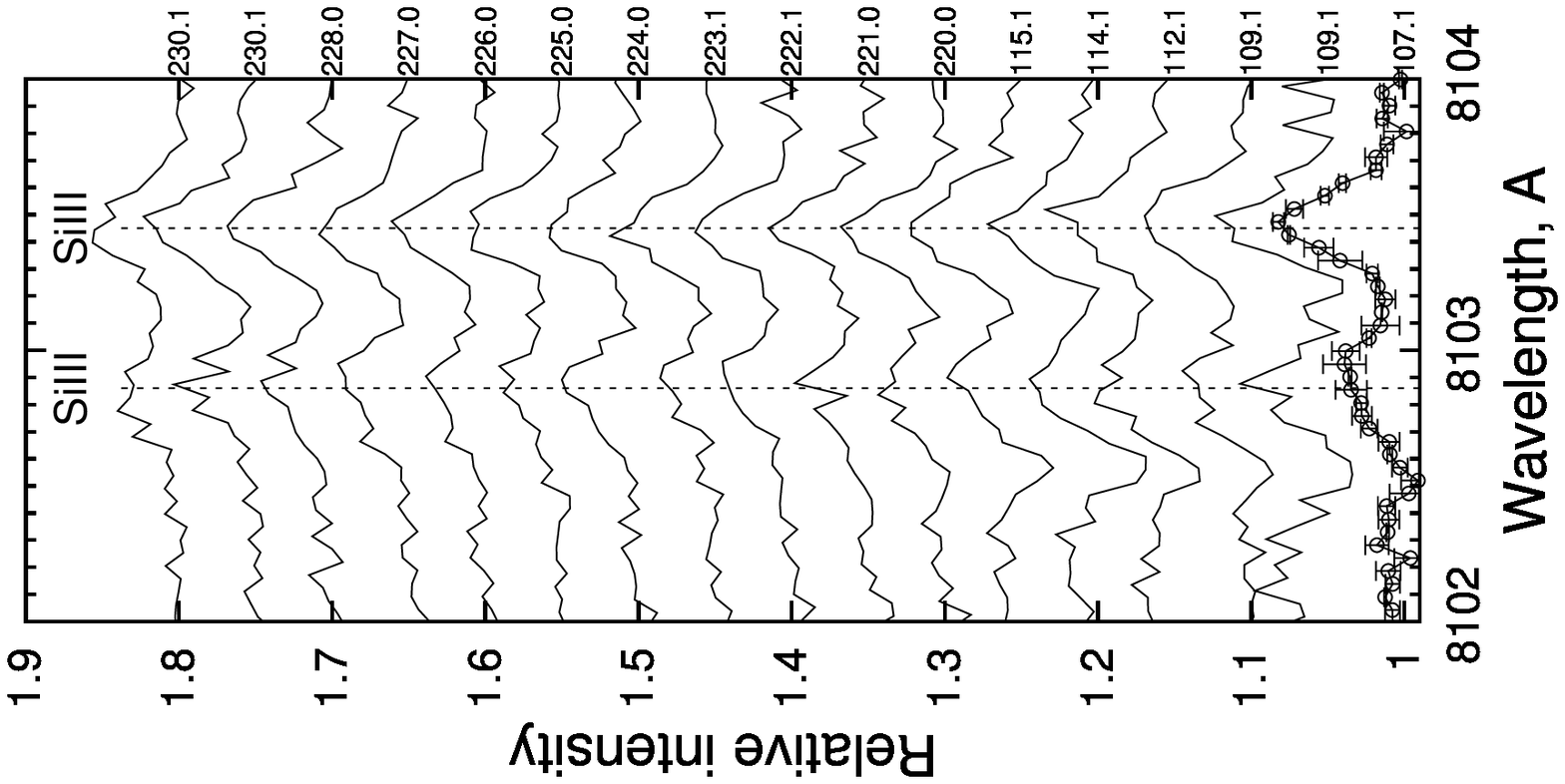}
\includegraphics[width=3.3in,angle=-90]{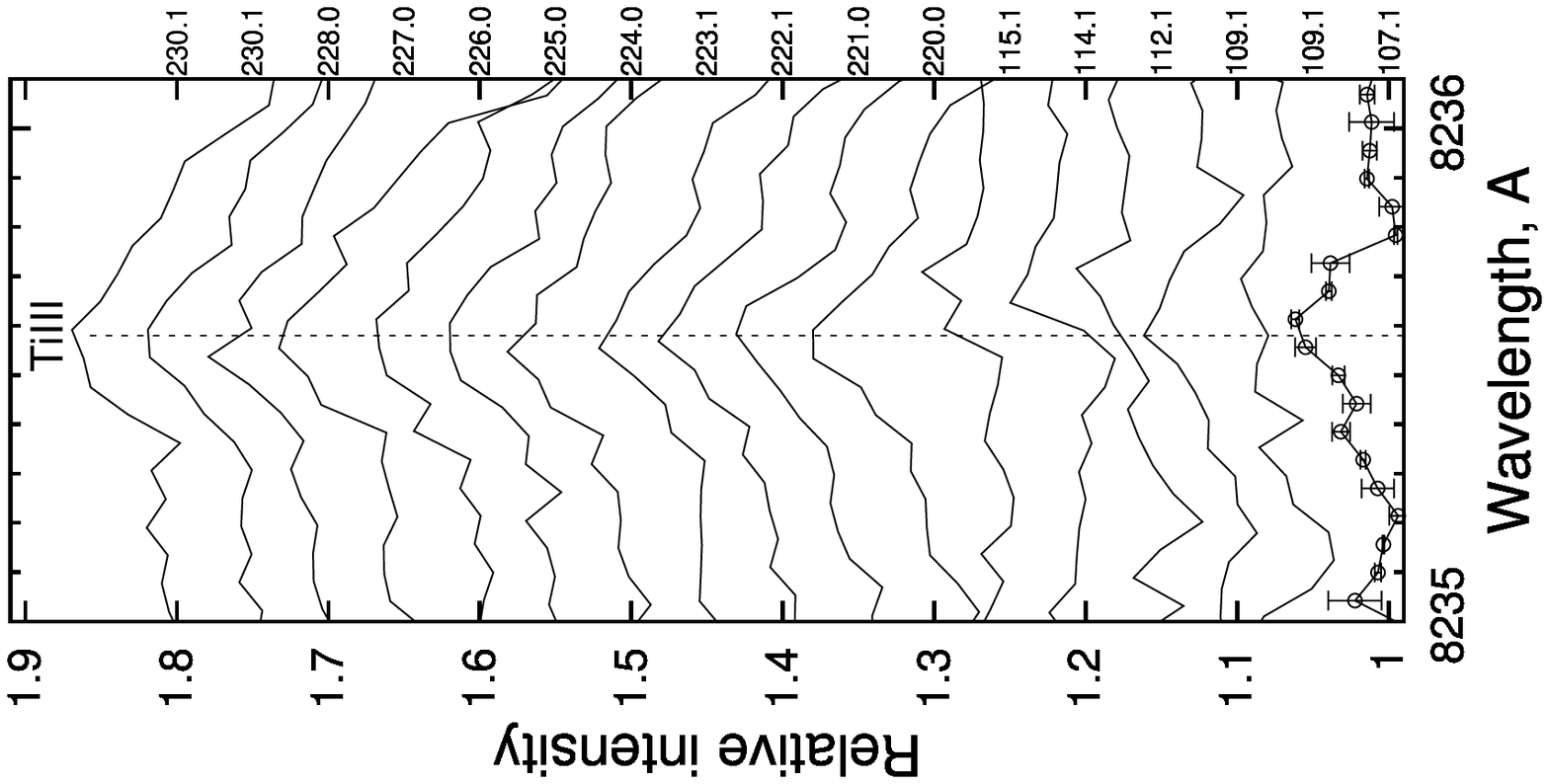}
\caption{ The same as Fig.~\ref{fig1}, but for Si\,{\sc iii} 8102.86\AA, 8103.45\AA\, (left panel) and Ti\,{\sc iii} 8235.58\AA\, (right panel) emission lines.}
\label{fig4}
\end{figure}

According to Ramspeck et al. \shortcite{ram01b}, HD~76431 has evolved past the HB phase and has fundamental parameters $T_{\rm eff}$ = 31000~K and log $g$ = 4.51.
In this article we have aimed to check those values and have obtained independent estimates of the atmospheric parameters of the target star
on the basis of our Steward Observatory spectra using the available grid of NLTE, fully metal-blanketed model atmospheres and synthetic spectra.
The fitting of Balmer, He\,{\sc i} and He\,{\sc ii} line profiles in the MMT and Bok spectra leads to results for $T_{\rm eff}$ and log $g$ that are similar
to the ones obtained by Ramspeck et al. \shortcite{ram01b}. We also have found that the NLTE effects are not important in the stellar atmosphere of HD~76431.

Ramspeck et al. \shortcite{ram01b} have also found that helium is underabundant by 0.5 dex and $V\sin{i}$ is less than 5 km s$^{-1}$. Our
estimate of the $\log{N_{\rm He}/N_{\rm H}}$ = $-$1.58$\pm$0.05 and $V\sin{i}$ = 3.5$\pm$0.5 km s$^{-1}$ are consistent with their results
and together with the underabundant helium, this suggests that diffusion could be active in this star.

The results of our abundance analysis presented in Table~{\ref{tab2} are in good accordance with previously published data of Ramspeck et al. \shortcite{ram01b}.
Nevertheless, we have not found the lines of P\,{\sc iii} reported by these authors in the spectra used here, but detected
several lines of Ar\,{\sc ii} that show a higher average abundance than its solar value.
We have also identified a few lines of Ti\,{\sc iii} 
that appears to be strongly overabundant (see Table~\ref{tab2}).
Also, the Si\,{\sc iii} 8102.86\AA, 8103.45\AA\, and Ti\,{\sc iii} 8235.58\AA\, lines are in emission
and probably originate from the upper layers of the stellar atmosphere of HD~76431(see. Fig.~\ref{fig4}).

The average abundance of carbon is close to its solar abundance, while nitrogen in average is slightly overabundant.
Our analysis suggests that carbon and nitrogen increase in concentration towards the deeper atmospheric layers.
Carbon and nitrogen become significantly overabundant in the deeper atmospheric layers, but nitrogen seems to
show a significant dispersion of abundance estimates obtained from the different spectral lines deeper in the atmosphere (see Fig.~\ref{fig2}).
The other chemical species (O\,{\sc ii} and Fe\,{\sc iii}) for which we have identified more than 10 lines do not show
signatures of vertical abundance stratification.
The use of stellar atmosphere models with higher and lower effective temperature ($\pm$2000~K) and the same gravity, and with higher and lower gravity
(log $g$ = 4.75 and 4.25) and the same effective temperature ($T_{\rm eff}$ = 31000~K) results in similar or larger vertical stratification of the carbon and nitrogen abundances. These additional tests make our conclusion that C and N are vertically stratified in HD~76431 more robust. This is the first detection of vertical abundance stratification in a post-HB star and it is the hottest star up to now to show such a feature.

\section{aknowledgements}
We thank the R\'{e}seau qu\'{e}b\'{e}cois de calcul de haute performance and Calcul Canada for computing resources.  This work was partially funded by the Natural Sciences and Engineering Research Council of Canada (NSERC) and la Facult\'{e} des \'{e}tudes sup\'{e}rieures et de la recherche de l'Universit\'{e} de Moncton.

\end{document}